\documentclass[reprint,NumberedRefs]{JASA}
\usepackage{multirow}
\usepackage{subfigure}

\begin{document}

\title[]{Two Chebyshev Spectral Methods for Solving Normal Modes in Atmospheric Acoustics}

\author{Tu Houwang}
\email{tuhouwang96@163.com}
\author{Wang Yongxian}
\email{yxwang@nudt.edu.cn}
\author{Xiao Wenbin}
\author{Lan Qiang}
\author{Liu Wei}

\affiliation{College of Meteorology and Oceanography, National University of Defense Technology, Changsha, 410073, China}

\preprint{H. Tu et al., JASA}		

\date{\today} 

\begin{abstract}
The normal mode model is important in computational atmospheric acoustics. It is often used to compute the atmospheric acoustic field under a harmonic point source. Its solution consists of a set of discrete modes radiating into the upper atmosphere, usually related to the continuous spectrum. In this article, we present two spectral methods, the Chebyshev--Tau and Chebyshev--Collocation methods, to solve for the atmospheric acoustic normal modes, and corresponding programs were developed. The two spectral methods successfully transform the problem of searching for the modal wavenumbers in the complex plane into a simple dense matrix eigenvalue problem by projecting the governing equation onto a set of orthogonal bases, which can be easily solved through linear algebra methods. After obtaining the eigenvalues and eigenvectors, the horizontal wavenumbers and their corresponding modes can be obtained with simple processing. Numerical experiments were examined for both downwind and upwind conditions to verify the effectiveness of the methods. The running time data indicated that both spectral methods proposed in this article are faster than the Legendre--Galerkin spectral method proposed previously.
\end{abstract}

\maketitle
\noindent \textbf{Keywords:} Chebyshev polynomial; normal modes; Tau method; collocation method; computational atmospheric acoustics.

\section{Introduction}
The propagation of sound waves in the atmosphere is a basic subject of atmospheric acoustics \cite{Salomons2001}. Sound waves in the atmosphere undergo a series of complex processes, including ground reflection, atmospheric scattering, refraction, and absorption \cite{Yang2015}. In fact, the propagation of sound waves in the atmosphere satisfies the wave equation, but it is difficult to strictly solve the wave equation. Thus, scientists make approximations to the wave equation for specific situations, thereby obtaining easy-to-solve equations, which can be solved numerically to obtain a solution of the sound field. Numerical sound fields have the advantages of intuitiveness and clarity, and they are widely used in acoustic research. Based on this idea of solving the numerical sound field, computational atmospheric acoustics, a sub-discipline of atmospheric acoustics, has been developed. Numerical models have many forms. Different models are suitable for different environments, and the results are not exactly the same. Mainstream numerical models include the parabolic equation (PE) \cite{Gilbert1989,Gilbert1993} model, the wavenumber integration method (the fast field program (FFP)) \cite{Finn2011,Raspet1985,Scooter2010}, and ray \cite{Pierce1991} and Gauss beam \cite{Yang2015} approaches. The normal mode model is also a fundamental method for solving for the acoustic field in the atmosphere with a finite ground impedance and horizontally stratified sound speed \cite{Yang2015,Finn2011,Salomons2001}. A horizontally stratified atmosphere allows the wave equation to be solved by the separation of variables method. After using Hankel integral transforms, the sound field can be expressed in terms of the sum of normal modes. When the ground impedance is complex or there is sound attenuation in the atmosphere, it is complicated to use the finite difference method to solve for the atmospheric normal modes, and the result is not very accurate \cite{Yang2015}.

In recent years, progress has been made on using spectral methods to solve underwater acoustic problems, and small-scale research has begun to link the spectral methods with the normal modes of underwater acoustics. Dzieciuch \cite{Dzieciuch1993} developed MATLAB code for computing normal modes based on Chebyshev approximations. Although he only realized the calculation of the simple Munk waveguide, this was the first step in the application of the spectral methods to computational ocean acoustics. In 2016, Evans \cite{Evans2016} used the Legendre--Galerkin spectral method to develop a sound propagation calculation program in a layered ocean environment. Subsequently, Tu et al. \cite{Tu2020a,Tu2020b,Tu2020bcode} used the Chebyshev--Tau spectral method to develop a program for calculating sound propagation in single-layer and layered ocean environments. They subsequently solved for the normal modes in underwater acoustics using the Chebyshev--Collocation method and proved that both of the spectral methods had high accuracy \cite{Tu2020c}. They also applied the spectral methods to the parabolic approximation of underwater acoustics \cite{Tu2020a,Tu2020d,Tu2020e}. The results of these studies indicated that it is feasible to apply spectral methods for the calculation of underwater acoustics, and in many cases, it has higher accuracy than the finite difference method. Monographs on spectral methods have also confirmed this \cite{Gottlieb1977,Boyd2001,Canuto2006,Jieshen2011}. Throughout the history of the development of atmospheric acoustics, many methods in underwater acoustics have been introduced \cite{Salomons2001,Yang2015}. In computational atmospheric acoustics, spectral methods are rarely used to calculate the numerical sound field. In 2017, Evans \cite{Evans2017} successfully introduced the Legendre--Galerkin spectral method to construct atmospheric acoustic normal modes. He then further improved the method \cite{Evans2018} and proved the convergence of the method \cite{Evans2020}. 

In this article, we propose two spectral methods for calculating atmospheric acoustic normal modes. The results are compared with Evans's code  \cite{Evans2018}, the correctness of the two spectral methods proposed in this article was verified, and computational speeds of the two spectral methods were demonstrated. The text is organized as follows. Section 2 describes normal modes in the atmosphere mathematically. Section 3 provides brief descriptions of the Chebyshev--Tau and Chebyshev--Collocation spectral methods and introduces the discretization of atmospheric acoustic normal modes. In Section 4, two numerical experiments are shown to verify the correctness of the methods proposed in this article. Section 5 analyzes the running speed of the spectral methods, and Section 6 concludes this article.

\section{Atmospheric Normal Modes}
Acoustic theory reveals that the core of solving the acoustic field with a time-independent harmonic point source is the following wave equation \cite{Finn2011}:
\begin{equation}
\label{eq:1}
    \rho \nabla \cdot \left(\frac{1}{\rho}\nabla p \right)+k^2 p=0.
\end{equation}
In the above homogeneous Helmholtz equation, $\rho$ is the density of the media, $p$ is the sound pressure in the frequency domain to be solved, and $k$ represents the wavenumber, which is related to the frequency of the source and the spatial position. $\rho$, $p$, and $k$ are all functions of the spatial position, i.e., $\rho(x,y,z)$, $p(x,y,z)$, and $k(x,y,z)$, respectively. 

\begin{figure}[htbp]
\includegraphics[width=\linewidth]{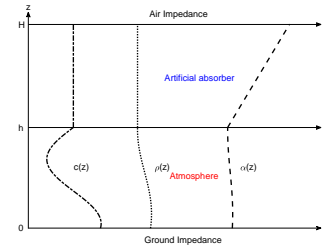}
\caption{\label{fig1} Atmospheric sound propagation environment.}
\end{figure}

We consider the medium of sound propagation to be the atmosphere depicted in \autoref{fig1}.
Taking the $\nabla$ operator in Eq.~\eqref{eq:1} in cylindrical coordinates to obtain the acoustic governing equation in the cylindrical coordinate system $(r, z)$, where $r$ is the range, and $z$ is the depth. Considering the case in \autoref{fig1} where the density $\rho(z)$ and wavenumber $k(z)$ are only related to depth (range-independent), Eq.~\eqref{eq:1} becomes:
\begin{equation}
\label{eq:2}
    \frac{1}{r}\frac{\partial}{\partial r}\left(
        r\frac{\partial p}{\partial r}
    \right) + 
    \rho (z) \frac{\partial}{\partial z}\left(
        \frac{1}{\rho (z)}\frac{\partial p}{\partial z}
    \right) +k^2(z)=0,
\end{equation}
where $k(z)=\omega /c(z)$, $\omega=2\pi f$ is the angular frequency of the sound source, $f$ is the frequency of the source, and $c(z)$ is the sound speed profile. When considering the attenuation of sound waves by the atmosphere, $k(z)=[1+i\eta\beta(z)]\omega/c(z)$, where $\beta(z)$ is the attenuating coefficient in units of dB per wavelength, and $\eta=(40\pi \log_{10}{e})^{-1}$. Through separation of variables, the acoustic pressure $p(r,z)$ can be decomposed as follows:
\begin{equation}
\label{eq:3}
    p(r,z)=\psi(z)R(r),
\end{equation}
where $R(r)$ can be approximated by an analytical form of the function, and $\psi (z)$ satisfies the following modal equation:
\begin{equation}
\label{eq:4}
    \rho(z)\frac{\mathrm{d}}{\mathrm {d}z}\left(
        \frac{1}{\rho(z)}\frac{\mathrm{d}\psi(z)}{\mathrm {d}z}
    \right) + k^2(z)\psi(z) = k_r^2 \psi(z).
\end{equation}
The modal equation is a Sturm--Liouville equation, and its characteristics are well known, that is, after adding appropriate boundary conditions, it has a series of modal solutions $(k_r,\psi)$, where $k_r^2$ is a constant. When the considered medium has attenuation, $k(z)$ is a complex function. The lower boundary of the atmosphere is the ground, and sound waves on the ground usually need to meet the following impedance boundary conditions:
\begin{equation}
\label{eq:5}
    \frac{\mathrm{d}\psi(0)}{\mathrm {d}z}+G\psi(0)=0,\quad G=ik(0)/Z,
\end{equation}
where $Z$ is the normalized ground impedance. The upper boundary of the atmosphere can be regarded as a free boundary at infinity, or it can be called an acoustic half-space condition. To make the problem finite and solvable via spectral methods, we add an artificial absorber layer $[h,H]$ above the interest area $[0, h]$, where the acoustic parameters in $[h,H]$ and $[0,h]$ must be continuous. The absorber layer is usually set to be thick enough to attenuate the sound energy propagating upward, and no energy is reflected back to the area of $[0,h]$. In this way, the following air impedance condition should be satisfied at $z=H$:
\begin{equation}
\label{eq:6}
    \frac{\mathrm{d}\psi(H)}{\mathrm {d}z}+\alpha\psi(H)=0,\quad \alpha=-ik(H).
\end{equation}
Solving the standard Sturm--Liouville problem will yield multiple sets of solutions $(k_{r_m},\psi_m), m=1,2,\cdots$, where $k_{r_m}$ is called the horizontal wavenumber, and $\psi_m$ is called the eigenmode or mode. The modes of Eq.~\eqref{eq:4} are arbitrary up to a nonzero scaling constant, so they should be normalized \cite{Finn2011} as follows:
\begin{equation}
\label{eq:7}
    \int_{0}^{H} \frac{{[\psi_m(z)}]^2}{\rho(z)}\mathrm {d} z=1,
    \quad m = 1, 2, \dots.
\end{equation}
Finally, the fundamental solution to the acoustic governing equation \eqref{eq:2} in the atmosphere can be approximated as follows \cite{Evans2017}:
\begin{equation}
\label{eq:8}
p(r,z) \approx \sqrt{2\pi} \sum_{m=1}^{M}\psi_m(z_s)\psi_m(z)\frac{\exp(k_{r_m}r)}{\sqrt{k_{r_m}r}},
\end{equation}
where $M$ is the number of modes used to synthesize the sound field.

The core of solving for the normal modes of atmospheric acoustics is the solution of the differential equations in Eq.~\eqref{eq:4}--\eqref{eq:6}. Solving for the normal modes of the atmospheric acoustics requires the discretization of Eq.~\eqref{eq:4}--\eqref{eq:6}. Traditionally, the domain of the problem solved by the spectral method is usually in the interval $[-1,1]$, so we first use $x=2z/H-1$ to scale the domain $z\in[0,H]\mapsto x\in[-1,1]$. Noting that $\mathrm {d} z/\mathrm {d} x=H/2$, we let the operators $\mathcal{L},\mathcal{P},\mathcal{Q}$ have the following forms:
\[
    \mathcal{L}=\left[\frac{4}{H^2}\rho(x)\frac{\mathrm{d}}{\mathrm {d}x}\left(\frac{1}{\rho(x)}\frac{\mathrm{d}}{\mathrm {d}x}
    \right) + k^2(x) \right],
\]
\[
    \mathcal{P}=\left[\frac{2}{H}\frac{\mathrm{d}}{\mathrm {d}x}+G\right],\quad \mathcal{Q}=\left[\frac{2}{H}\frac{\mathrm{d}}{\mathrm {d}x}+\alpha\right].
\]
Eq.~\eqref{eq:4}--\eqref{eq:6} can be written in the following form:
\begin{equation}
\begin{split}
\label{eq:9}
    &\mathcal{L}\psi(x)=k_r^2\psi(x),\quad x\in (-1,1),\\
    &\mathcal{P} \psi(-1)=0,\quad
    \mathcal{Q} \psi(1)=0.
\end{split}    
\end{equation} 
Next, we will develop two spectral methods to solve this system.

\section{Discretized Atmospheric Normal Modes by Two Spectral Methods}
A spectral method is a kind of weighted residual method, and it can provide accurate solutions to differential equations \cite{Boyd2001,Canuto2006}. In the spectral method, the unknown function to be solved $\psi(x)$ is expanded by a set of linearly independent bases $\phi_k(x)$. When the number of bases tends to infinity, an accurate representation of $\psi(x)$ can be obtained. However, in actual calculations, it is usually necessary to truncate to the first $N$-order terms, thus obtaining an approximation of $\psi(x)$, as follows:
\begin{equation}
\label{eq:10}
   \psi(x)=\sum_{k=0}^{\infty}\hat{\psi}_k\phi_k(x) \approx \psi_N(x)= \sum_{k=0}^{N} \hat{\psi}_k\phi_k(x),
\end{equation}
where $\hat{\psi}_k$ represents the expansion coefficients. Obtaining the value of $\hat{\psi}_k$ is equivalent to obtaining the approximate solution $\psi_N(x)$ of $\psi(x)$. Inserting $\psi_N(x)$ from Eq.~\eqref{eq:10} into Eq.~\eqref{eq:9}, Eq.~\eqref{eq:9} is no longer strictly true, and there is a residual $Res(x)$, defined as follows:
\begin{equation}
\label{eq:11}
   Res(x)=\mathcal{L}\psi_N(x)-k_r^2\psi_N(x).
\end{equation}
To make $\psi_N(x)$ as close to $\psi(x)$ as possible, we need to minimize the residual through a certain principle \cite{Jieshen2011}. Setting the weighted integral of the residuals equal to zero is a widely used principle \cite{Boyd2001}:
\begin{equation}
\label{eq:12}
   \int_{-1}^1 Res(x)w(x) \mathrm {d}x=0.
\end{equation}
From Eq.~\eqref{eq:9}, the residual $Res(x)$ can be minimized only by adjusting the value of the expansion coefficients $\hat{\psi}_k$. The choice of the weight function $w(x)$ is also crucial. In the two spectral methods developed in this article, the basis functions $\phi_k(x)$ are both Chebyshev polynomials $T_k(x)$, and the difference is the selection of weight functions. The Chebyshev polynomial basis functions are provided in the Appendix of this article.

\subsection{Discretized Atmospheric Normal Modes by Chebyshev--Tau Spectral Method}
In the Chebyshev--Tau spectral method, in addition to the basis functions being Chebyshev polynomials ($\phi_k(x)=T_k(x)$), the weight functions are also Chebyshev polynomials ($w(x)=T_k(x)$). Inserting Eq.~\eqref{eq:10} and \eqref{eq:11} into Eq.~\eqref{eq:12}, we obtain the new form of Eq.~\eqref{eq:12} for the Chebyshev--Tau spectral method:
\begin{equation}
\begin{split}
	&\int_{-1}^1 \frac{T_j(x)}{\sqrt{1-x^2}}\left(\mathcal{L} \sum_{k=0}^{N}\hat{\psi}_k T_k(x)-k_r^2\sum_{k=0}^{N}\hat{\psi}_k T_k(x)\right) \mathrm {d}x =0\\
	&j=0,1,\cdots,N-2,..
\end{split}
\label{eq:13}
\end{equation}
where $\frac{1}{\sqrt{1-x^2}}$ is the orthogonal weighting factor of the Chebyshev polynomial basis function space.

This equation is also known as the weak form of Eq.~\eqref{eq:9}. It will form $(N-1)$ algebraic equations (excluding the boundaries of $x$), the two boundary conditions will produce two algebraic equations, and the unknowns to be solved for are $\hat{\psi}_0$ to $\hat{\psi}_N$. The integral formulas listed in the above equations can be computed by the Gauss--Lobatto quadrature \cite{Canuto2006} to obtain accurate results. To include the two end points of the domain $x$, the Gauss--Lobatto nodes on $x\in[-1,1]$ are taken \cite{Canuto2006}:
\begin{equation}
\label{eq:14}
    x_j=-\cos \left(\frac{j \pi}{N} \right),\quad j=0,1,\cdots,N.
\end{equation}
There are two forms of Gauss--Lobatto quadrature  \cite{Canuto2006}:
\begin{equation}
\label{eq:15}
\begin{split}
    &\int_{-1}^1 \frac{f(x)}{\sqrt{1-x^2}}\mathrm {d}x \approx \sum_{j=0}^{N} f(x_j) \omega_j,\\
    &\int_{-1}^1 f(x)\mathrm {d}x \approx \sum_{j=0}^{N} f(x_j)\omega_j \sqrt{1-x_j^2},\\ 
    &\omega_{j}=\left\{\begin{array}{ll}
    \frac{\pi}{2 N}, & j=0, N \\
    \frac{\pi}{N}, & \text { otherwise }
    \end{array}\right.,
\end{split}
\end{equation}
where $f(x)$ is the function to be integrated.

We convert the original solution of the unknown function $\psi(x)$ into solving for its expansion coefficients $\hat{\psi}_k$ under the Chebyshev polynomial basis. The only difficulty is the discretization of the operator $\mathcal{L}$. The conclusions used in the following text are directly given here. For detailed derivations, readers can refer to Refs.~\onlinecite{Gottlieb1977,Canuto2006,Boyd2001,Jieshen2011}.

The derivative $\frac{\mathrm {d}}{\mathrm {d}x}$ is included in the $\mathcal{L}$ operator, the expanded coefficients $\hat{\psi}^{\prime}_{k}$ of $\psi^{\prime}(x)$ satisfy the following relationship with $\hat{\psi}_{k}$:
\begin{equation}
\label{eq:16}
    \hat{\psi}_{k}^{\prime} \approx \frac{2}{c_{k}} 
    \sum_{\substack{j=k+1,\\ 
				j+k=\mathrm{odd}}}^{N} j \hat{\psi}_{j},\quad
	\bm{\hat{\psi}'}=\mathbf{\hat{D}}\bm{\hat{\psi}},
\end{equation}
the second formula is the vector form of the first formula, where column vectors $\bm{\hat{\psi}}=[\hat{\psi}_0,\cdots,\hat{\psi}_N]^T$ and $\bm{\hat{\psi}'}=[\hat{\psi}'_0,\cdots,\hat{\psi}'_N]^T$, respectively. $\mathbf{\hat{D}}$ is a square matrix of order $(N+1)$. To distinguish it from the differential matrix $\mathbf{D}$ in the Chebyshev--Collocation spectral method, a hat symbol is added to the relationship matrix.

The known function $k^2(x)$ is included in the $\mathcal{L}$ operator. Letting $v(x)=k^2(x)$, there will be a product term $y(x)=v(x)\psi(x)$, and the expanded coefficients $\hat{y}_{k}$ of $y(x)$ satisfy the following relationship with $\hat{\psi}_{k}$:
\begin{equation}
	\label{eq:17}
	\hat{y}_k \approx 
	\frac{1}{2} \sum_{m+n=k}^{N} \hat{\psi}_m\hat{v}_n +
	\frac{1}{2} \sum_{|m-n|=k}^{N} \hat{\psi}_m\hat{v}_n,\quad \mathbf{\hat{y}} \approx \mathbf{\hat{C}}_{v} \bm{\hat{\psi}}. 
\end{equation}
Similarly, the second formula is the equivalent vector form, $\mathbf{\hat{C}}_{v}$ is also a square matrix of order $(N+1)$, and the subscript $v$ indicates that the known function in the operator is $v(x)$.

We show the discretization of the first term of the operator $\mathcal{L}$. We let
\begin{equation}
\label{eq:18}
    g(x)= \frac{1}{\rho(x)}, 
    \quad
    s(x)= g(x) \psi'(x), 
    \quad
    f(x)= \rho(x) s'(x).
\end{equation}
In the Chebyshev--Tau spectral method, applying Eqs.~\eqref{eq:16} and \eqref{eq:17} to Eq.~\eqref{eq:18}, we can obtain
\begin{equation}
\label{eq:19}
    \mathbf{\hat{s}}= \mathbf{\hat{C}}_g(\mathbf{\hat{D}}\bm{\hat{\psi})},\quad
    \mathbf{\hat{f}}= \mathbf{\hat{C}}_{\rho}(\mathbf{\hat{D}}\mathbf{\hat{s}})= \mathbf{\hat{C}}_{\rho}(\mathbf{\hat{D}}(\mathbf{\hat{C}}_g(\mathbf{\hat{D}}\bm{\hat{\psi}}))).
\end{equation}
The discrete forms of the operator $\mathcal{L}$ and Eq.~\eqref{eq:9} are as follows:
\begin{equation}
\label{eq:20}
    \mathbf{\hat{L}}=\left(\frac{4}{H^2}\mathbf{\hat{C}}_{\rho}\mathbf{\hat{D}}\mathbf{\hat{C}}_g\mathbf{\hat{D}}+\mathbf{\hat{C}}_{v} \right),\quad \mathbf{\hat{L}}\bm{\hat{\psi}}=k_r^2\bm{\hat{\psi}}.    
\end{equation}

The boundary conditions produce algebraic equations about the expansion coefficients in the Chebyshev--Tau spectral method as follows. In the Tau method, the function $\psi(x= \pm 1)$ in the boundary conditions is also expanded by Eq.~\eqref{eq:10}. The discretization of the boundary operators $\mathcal{P}$ and $\mathcal{Q}$ is similar to that of operator $\mathcal{L}$, so the two boundary conditions generate two equations related to $\hat{\psi}_k$. To facilitate the description of the processing of the boundary conditions, the following intermediate row vectors are defined as follows:
\[
\mathbf{\hat{t}}_1=[T_0(-1),T_1(-1),\cdots,T_N(-1)],\quad
\mathbf{\hat{p}}=\frac{2}{H}\mathbf{\hat{t}}_1\mathbf{\hat{D}}+G\mathbf{\hat{t}}_1,
\]
\[
\mathbf{\hat{t}}_2=[T_0(1),T_1(1),\cdots,T_N(1)],\quad
\mathbf{\hat{q}}=\frac{2}{H}\mathbf{\hat{t}}_2\mathbf{\hat{D}}+\alpha\mathbf{\hat{t}}_2.
\]
The matrix form of the discrete ground and air boundary conditions Eq.~\eqref{eq:5} and \eqref{eq:6} in the Chebyshev--Tau spectral method can be written as follows:
\begin{equation}
\label{eq:21}
    \mathbf{\hat{p}}\bm{\hat{\psi}}=0,\quad  \mathbf{\hat{q}}\bm{\hat{\psi}}=0. 
\end{equation}
The algebraic equations formed by these two boundary conditions and the $(N-1)$ algebraic equations obtained from the weak form are solved simultaneously, and  we can then solve for $\hat{\psi}_k$ and obtain $\psi(x)$.

The row vectors $\mathbf{\hat{p}}$ and $\mathbf{\hat{q}}$ are used to replace the last two rows of the $\mathbf{\hat{L}}$ matrix in Eq.~\eqref{eq:20}, and the last two elements of the column vector $\bm{\hat{\psi}}$ on the right-hand side of Eq.~\eqref{eq:20} are replaced with 0, so that the boundary conditions are strictly met. We let the matrix composed of the first $(N-1)$ rows and $(N-1)$ columns of $\mathbf{\hat{L}}$ be $\mathbf{\hat{L}}_{11}$. The matrix composed of the first $(N-1)$ rows and the last two columns of $\mathbf{\hat{L}}$ is $\mathbf{\hat{L}}_{12}$. The row vectors composed of the first $(N-1)$ elements of the row vectors $\mathbf{\hat{p}}$, $\mathbf{\hat{q}}$, and $\bm{\hat{\psi}}$ are $\mathbf{\hat{p}}_1$, $\mathbf{\hat{q}}_1$ and $\bm{\hat{\psi}}_1$, respectively. The row vectors composed of the last two elements of the row vectors $\mathbf{\hat{p}}$, $\mathbf{\hat{q}}$, and $\bm{\hat{\psi}}$ are $\mathbf{\hat{p}}_2$, $\mathbf{\hat{q}}_2$, and $\bm{\hat{\psi}}_2$, respectively. Thus, Eq.~\eqref{eq:20} can be changed to the following block form:
\begin{equation}
\label{eq:22}
\left[
\begin{array}{c|c}
\mathbf{\hat{L}}_{11}&\mathbf{\hat{L}}_{12}\\
\hline
\mathbf{\hat{p}}_1&\mathbf{\hat{p}}_2\\
\mathbf{\hat{q}}_1&\mathbf{\hat{q}}_2
\end{array}
\right]\left[
\begin{array}{c}
\bm{\hat{\psi}}_1\\
\hline
\hat{\psi}_{N-1}\\
\hat{\psi}_N
\end{array}
\right]=k_r^2\left[
\begin{array}{c}
\bm{\hat{\psi}}_1\\
\hline
0\\
0
\end{array}\right].
\end{equation}
According to the horizontal and vertical lines in the above formula, Eq.~\eqref{eq:22} can be abbreviated as follows:
\begin{equation}
\label{eq:23}
\left[
\begin{array}{cc}
\mathbf{\hat{L}}_{11}&\mathbf{\hat{L}}_{12}\\
\mathbf{\hat{L}}_{21}&\mathbf{\hat{L}}_{22}
\end{array}
\right]\left[
\begin{array}{c}
\bm{\hat{\psi}}_1\\
\bm{\hat{\psi}}_2
\end{array}
\right]=k_r^2\left[
\begin{array}{c}
\bm{\hat{\psi}}_1\\
\mathbf{0}
\end{array}\right].
\end{equation}
Eq.~\eqref{eq:23} can be solved as follows:
\begin{equation}
\label{eq:24}
    \bm{\hat{\psi}}_2=-\mathbf{\hat{L}}_{22}^{-1}\mathbf{\hat{L}}_{21}\bm{\hat{\psi}}_1,\quad(\mathbf{\hat{L}}_{11}-\mathbf{\hat{L}}_{12}\mathbf{\hat{L}}_{22}^{-1}\mathbf{\hat{L}}_{21})\bm{\hat{\psi}}_1=k_r^2\bm{\hat{\psi}}_1.
\end{equation}
Therefore, a set of $(k_r^2,\bm{\hat{\psi}})$ can be solved for by the $(N-1)$th-order matrix eigenvalue problem in Eq.~\eqref{eq:24}. For each set of eigenvalues/eigenvectors $(k_{r_m}^2,\bm{\hat{\psi}}_m)$, an eigensolution $(k_{r_m},\bm{\psi}_m)$ of Eq.~\eqref{eq:4} can be obtained by Eq.~\eqref{eq:10}. In this process, each eigenmode should be normalized by Eq.~\eqref{eq:7}. Finally, the sound pressure field is obtained by applying Eq.~\eqref{eq:8} to the chosen modes.

\subsection{Discretized Atmospheric Normal Modes by Chebyshev--Collocation Spectral Method}

The Collocation method uses the Dirac function $\delta(x)$ as the weight function in Eq.~\eqref{eq:12}. The characteristics of the $\delta(x)$ function are well known. In the Collocation method, Eq.~\eqref{eq:12} becomes the following:
\begin{equation}
\label{eq:25}
\begin{split}
    &\int_{-1}^{1} Res(x) \delta(x-x_j) \mathrm {d}x
	= Ri(x_j)=\mathcal{L}\psi(x_j)-k_r^2\psi(x_j),\\
	&\quad j=0,1,2,...N.
\end{split}
\end{equation}
The above formula shows that in the Collocation method, the weighted residual principle becomes that the residuals are all 0 at the selected discrete points $x_j$. Its essence is to only make the original differential equation \eqref{eq:9} strictly hold on this set of discrete points, so as to solve for the function value $\psi(x_j)$ of the modal function $\psi(x)$ on this set of discrete points as an approximation. In the Collocation method, there is no need to expand the function to be sought as Eq.~\eqref{eq:10}. This is why the Collocation method is considered to be a special spectral method, sometimes called the pseudospectral method \cite{Canuto2006}. In the Chebyshev--Collocation method, we also take the discrete points of the Chebyshev--Gauss--Lobatto nodes in Eq.~\eqref{eq:14}. In this case, the only difficulty is the discretization of operator $\mathcal{L}$. The conclusions used in the following text are directly given here as in the introduction of Chebyshev--Tau spectral method. For a detailed derivation, readers can refer to Refs.~\onlinecite{Gottlieb1977,Boyd2001,Canuto2006,Jieshen2011}.

The derivative term $\psi'(x)$ and $\psi(x)$ have the following relationship:
\begin{equation}
\label{eq:26}
\begin{split}
&\bm{\psi}'=\mathbf{D}\bm{\psi},\quad
\mathbf{D}=\left\{\begin{array}{ll}
\frac{c_{j}(-1)^{j+l}}{c_{k}\left(x_{j}-x_{k}\right)}, & j \neq k \\
\frac{-x_{k}}{2\left(1-x_{k}^{2}\right)}, & 1 \leq j=k \leq N-1 \\
\frac{2 N^{2}+1}{6}, & j=l=0 \\
-\frac{2 N^{2}+1}{6}, & j=l=N
\end{array}\right.\\
&c_{k}=\left\{\begin{array}{ll}
	2, & k=0 \\
	1, & k>0 \end{array}\right.,
\end{split}
\end{equation}
where $\bm{\psi}'=[\psi'(x_0),\psi'(x_1), \psi'(x_2),\cdots, \psi'(x_N)]^T$ represents the function value of the derivative term $\psi'(x)$. Similarly, $\bm{\psi}=[\psi(x_0),\psi(x_1), \psi(x_2),\cdots, \psi(x_N)]^T$. Matrix $\mathbf{D}$ is also called the Chebyshev--Collocation differential matrix. 

The product $y(x)=v(x)\psi(x)$ can be processed as follows:
\begin{equation}
\label{eq:27}
\mathbf{y}=\mathbf{C}_v\bm{\psi},
\end{equation}
where $\mathbf{y}=[y(x_0),y(x_1), y(x_2),\cdots, y(x_N)]^T$, $\mathbf{C}_v$ is a $(N+1)\times(N+1)$ diagonal matrix, and $(\mathbf{C}_v)_{ii}=v(x_i),i=0,1,...,N$.

For the Collocation method, the boundary conditions are only related to the endpoints of the domain $x$, so the discrete points on the boundaries ($x_0$ and $x_N$) only need to satisfy the boundary conditions, not the differential equation. The discretized forms of the operators $\mathcal{P}$ and $\mathcal{Q}$ are similar to that of operator $\mathcal{L}$. Similar to the Chebyshev--Tau spectral method, the operator $\mathcal{L}$ also needs to be discretized in the Chebyshev--Collocation method. With reference to Eq.~\eqref{eq:18} and \eqref{eq:19}, in the Chebyshev--Collocation method, the $\mathcal{L}$ operator and Eq.~\eqref{eq:9} have the following forms:
\begin{equation}
\label{eq:28}
    \mathbf{L}=\left(\frac{4}{H^2}\mathbf{C}_{\rho}\mathbf{D}\mathbf{C}_g\mathbf{D}+\mathbf{C}_{v} \right),\quad \mathbf{L}\bm{\psi}=k_r^2\bm{\psi}.    
\end{equation}
To facilitate the description of the processing of the boundary conditions, the first and last rows of $\mathbf{D}$ are defined as row vectors $\mathbf{d}_1$ and $\mathbf{d}_2$, respectively. The following intermediate $(N+1)$-dimensional row vectors are defined as follows:
\[
\mathbf{t}_1=[1,0,\cdots,0],\quad
\mathbf{p}=\frac{2}{H}\mathbf{d}_1+G\mathbf{t}_1,
\]
\[
\mathbf{t}_2=[0,0,\cdots,1],\quad
\mathbf{q}=\frac{2}{H}\mathbf{d}_2+\alpha\mathbf{t}_2.
\]
The matrix form of the discrete ground and air impedance conditions Eq.~\eqref{eq:5} and \eqref{eq:6} in the Chebyshev--Collocation method can be written as follows:
\begin{equation}
\label{eq:29}
    \mathbf{p}\bm{\psi}=0,\quad \mathbf{q}\bm{\psi}=0.   
\end{equation}

In the Collocation method, the row vectors $\mathbf{p}$ and $\mathbf{q}$ are used to replace the first row and the last row of the $\mathbf{L}$ matrix in Eq.~\eqref{eq:28}, so that the boundary conditions are satisfied. We let the block matrix formed by the second row to the $N$-th row of the matrix $\mathbf{L}$ be $\mathbf{L}_{1}$, and the column formed by the second to the $N$-th elements of $\bm{\psi}$ be $\bm{\psi}_1$. Eq.~\eqref{eq:28} can then be written as follows:
\begin{equation}
\label{eq:30}
\left[
\begin{array}{c}
\mathbf{p}\\
\mathbf{L}_1\\
\mathbf{q}
\end{array}
\right]\left[
\begin{array}{c}
\psi_0\\
\bm{\psi}_1\\
\psi_N
\end{array}
\right]=k_r^2\left[
\begin{array}{c}
0\\
\bm{\psi}_1\\
0
\end{array}\right].
\end{equation}
We only need to perform a simple row transformation and column transformation on Eq.~\eqref{eq:30} to transform it to a form similar to Eq.~\eqref{eq:23}, and then we use the same method used for Eq.~\eqref{eq:24} to find the eigenvalues/eigenvectors. A set of $(k_r^2,\bm{\psi})$ can be solved by the $(N-1)$th-order matrix eigenvalue problem in Eq.~\eqref{eq:24}. Each eigenmode should be normalized by Eq.~\eqref{eq:7}. Finally, the sound pressure field is obtained by applying Eq.~\eqref{eq:8} to the chosen modes.

\section{Numerical experiment and analysis}
To verify the correctness of the above presented spectral methods in solving the normal modes of atmospheric acoustics, the authors developed the corresponding programs based on the above derivation. The programs based on Chebyshev--Tau spectral method and Chebyshev--Collocation method are called `AtmosCTSM' and `AtmosCCSM,' respectively. The code was written in FORTRAN/MATLAB and is available at the author's GitHub homepage (\url{https://github.com/tuhouwang/Atmospheric-normal-modes}). For comparison, we considered the program `aaLG' based on the Legendre--Galerkin spectral method, which was developed by Evans in FORTRAN and verified by comparison with PE and FFP \cite{Evans2018}. The two examples shown by Evans \cite{Evans2017} can be used as benchmark examples. The source frequency of both cases was 100 Hz at a height of 5 m above the ground. The normalized ground impedance $Z$ (related to the constant $G$ in Eq.~\eqref{eq:6}) is the same as the value used by Gilbert \cite{Gilbert1993}, and the value is $Z=12.97+12.38i$. In the following two experiments, the order of the spectral truncation $N$ in the three spectral methods was taken as 1500. Using the TL to express the acoustic field \cite{Finn2011}, the relationship between it and the sound pressure is  $\text{TL}=-20\log_{10}({|p|}/{|p_0|})$, where $p_0$ is the sound pressure at a distance of 1 m from the sound source.

\subsection{Downwind Case}
The first numerical experiment was a downwind case. The piecewise linear acoustic parameter profile used in this numerical experiment, which was presented by Evans \cite{Evans2017}, is shown in Table \ref{tab1}. In contrast to the case considered by Evans \cite{Evans2017,Evans2018}, the change of the atmospheric density with height was considered in this work. Therefore, a column of density data is included in the table. In fact, when the density is taken as a constant, $\rho(z)$ and $1/\rho(z)$ in Eq.~\eqref{eq:5} will be eliminated, which means that the uniform density has no effect on the propagation of normal modes. The table clearly reveals that the thickness of the atmosphere is 700 m, and the artificial absorber is located between 700 and 2000 m.

\begin{table}[htbp]
\tiny
\caption{\label{tab1} Piecewise linear acoustic parameter profile used in experiment 1, cited from Evans \cite{Evans2017}.}
\begin{ruledtabular}
\begin{tabular}{lccc}
Height (m)& Sound speed (m/s)& Attenuation (dB/wavelength)& Density (kg/m$^3$) \\
\hline
2000 & 344.0 & 2.50 & $\multirow{7}{*}{constant}$\\
1500 & 344.0 & 0.10 &  \\
900  & 344.0 & 0.01 &  \\
700  & 344.0 & 0.00 &  \\
500  & 341.5 & 0.00 &  \\
100  & 349.0 & 0.00 &  \\
0    & 345.0 & 0.00 &  \\
\end{tabular}
\end{ruledtabular}
\end{table}

\begin{figure}[htbp]
\includegraphics[width=8cm]{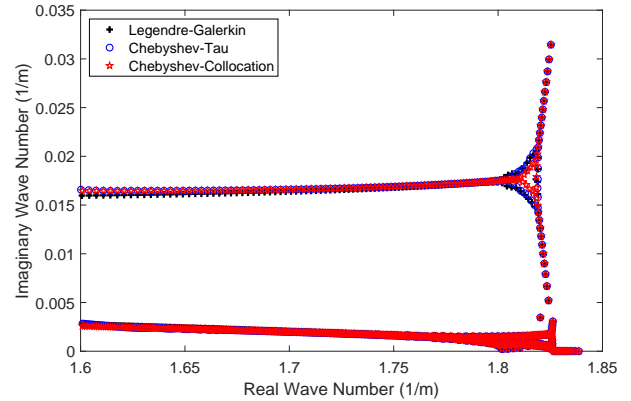}
\caption{\label{fig2} Horizontal wavenumbers calculated by the Legendre--Galerkin spectral method and the two spectral methods proposed in this article for experiment 1.}
\end{figure}

\begin{figure}[htbp]
\includegraphics[width=8cm]{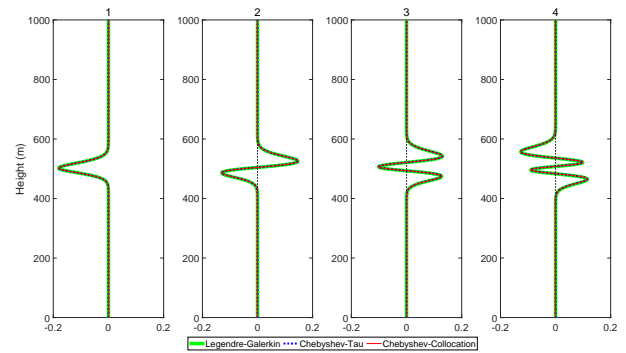}
\caption{\label{fig3} First four normal modes calculated by the three spectral methods in experiment 1.}
\end{figure}

\begin{figure}[htbp]
\subfigure[]{\includegraphics[width=8cm]{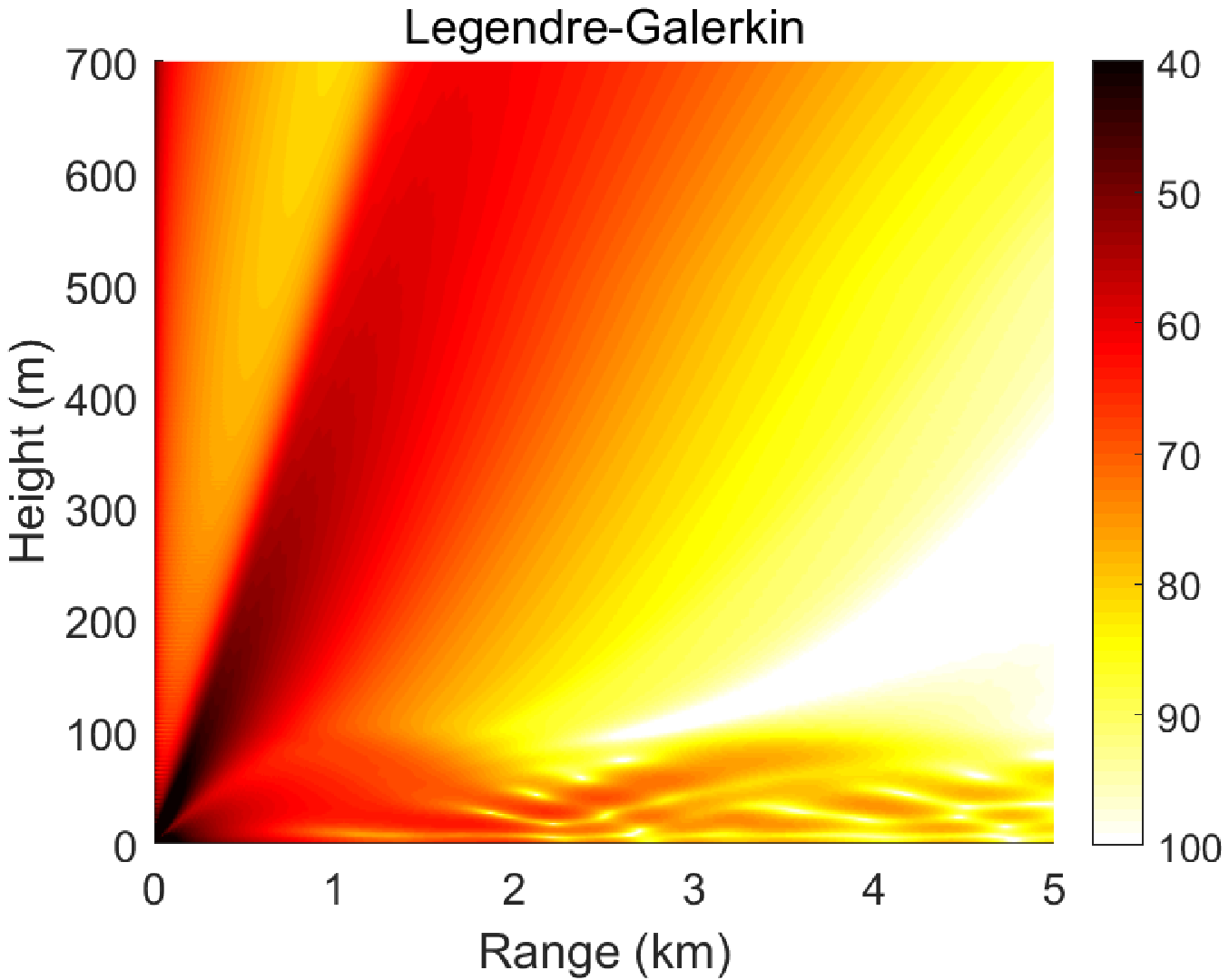}}
\subfigure[]{\includegraphics[width=8cm]{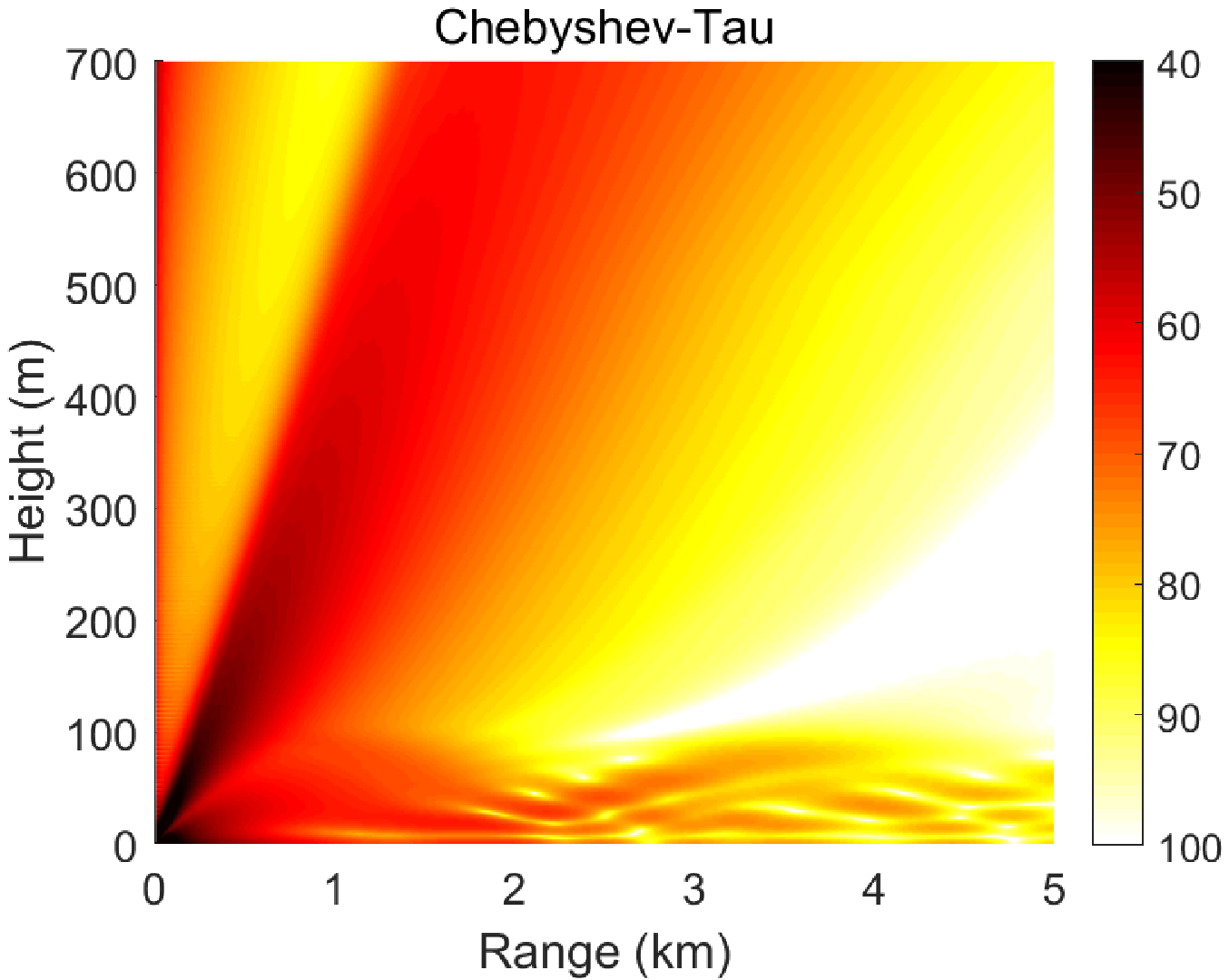}}
\subfigure[]{\includegraphics[width=8cm]{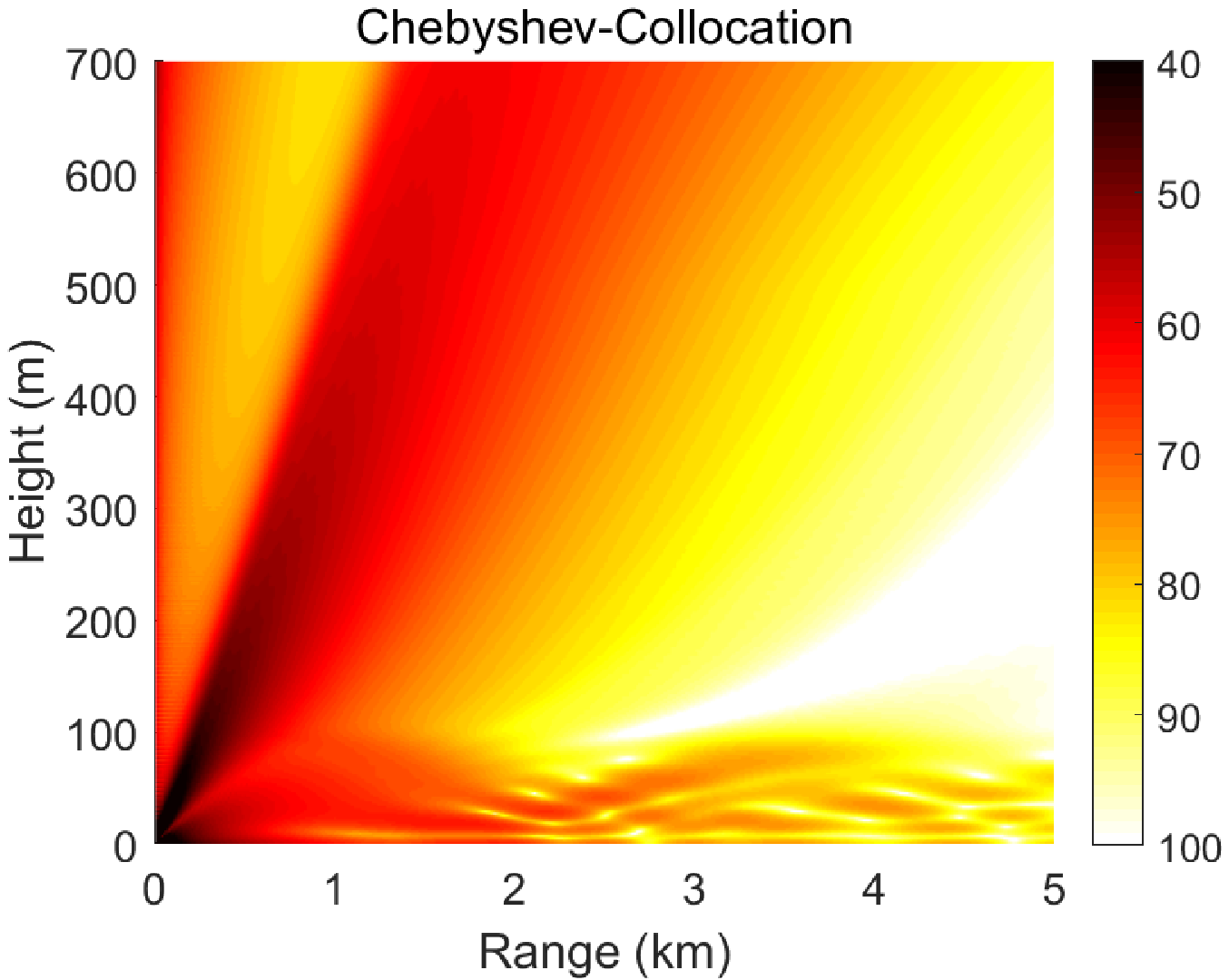}}
\caption{\label{fig4} Atmospheric acoustic fields obtained by (a) Legendre--Galerkin spectral method \cite{Evans2018}, (b) the Chebyshev--Tau spectral method, and (c) the Chebyshev--Collocation method of experiment 1.}
\end{figure}

\begin{figure}[htbp]
\includegraphics[width=8cm]{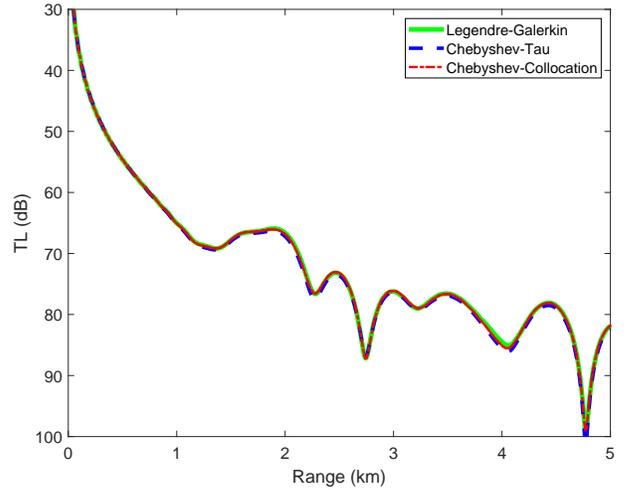}
\caption{\label{fig5} TL versus range for a receiver at a height of 1 m over the range interval 0--5 km from Legendre--Galerkin spectral method \cite{Evans2018}, Chebyshev--Tau spectral method, and Chebyshev--Collocation method.}
\end{figure}

\autoref{fig2} shows the horizontal wavenumbers $k_r$ calculated by the Legendre--Galerkin spectral method and the two spectral methods developed in this article on the complex plane. The consistency of the eigenvalue distribution in the figure illustrates the correctness of the horizontal wavenumbers calculated by the three methods. \autoref{fig3} shows the first four normal modes of experiment 1. It reveals that the modes obtained by the two spectral methods proposed in this article were highly consistent with those obtained by the Legendre--Galerkin spectral method. \autoref{fig4} presents an overview of the acoustic fields obtained by the three methods. We used the first 552 modes with phase velocities between 341.7 and 391.2 m/s to synthesize the sound fields. The horizontal wavenumbers of these modes are shown in \autoref{fig2}. The acoustic fields calculated by the three methods were very similar. \autoref{fig5} shows the TL curves versus the range for a receiver at a height of 1 m over the range interval 0--5 km. The results of the two spectral methods presented in this article were very similar to those of the Legendre--Galerkin spectral method, and there may have been small differences only in the acoustic shadow areas.

\subsection{Upwind Case}
The second numerical experiment is an upwind case. The piecewise linear acoustic parameter profile used in this numerical experiment was presented by Evans \cite{Evans2017}, and it is shown in Table \ref{tab2}. The table clearly illustrates that the thickness of the atmosphere was 900 m, and the artificial absorber was located between 900 and 2000 m. 

\begin{table}[htbp]
\tiny
\caption{\label{tab2} Piecewise linear acoustic parameter profile used in experiment 2, cited from Evans \cite{Evans2017}.}
\begin{ruledtabular}
\begin{tabular}{lccc}
Height (m)& Sound speed (m/s)& Attenuation (dB/wavelength)& Density (kg/m$^3$) \\
\hline
2000 & 346.0 & 1.00 & $\multirow{7}{*}{constant}$\\
1500 & 346.0 & 0.10 &  \\
1200 & 346.0 & 0.01 &  \\
900  & 346.0 & 0.00 &  \\
500  & 348.0 & 0.00 &  \\
350  & 344.0 & 0.00 &  \\
100  & 340.0 & 0.00 &  \\
0    & 344.0 & 0.00 &  \\
\end{tabular}
\end{ruledtabular}
\end{table}

\begin{figure}[htbp]
\includegraphics[width=8cm]{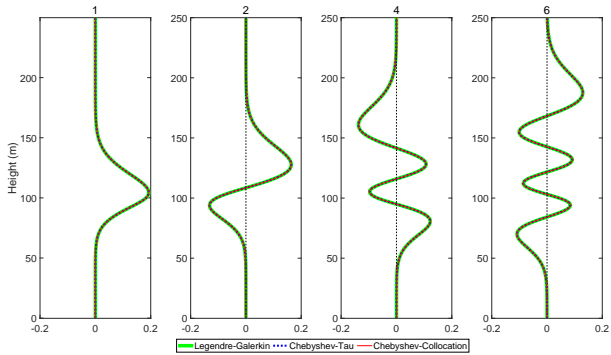}
\caption{\label{fig6} (a)--(d) First, second, fourth, and sixth modes obtained by Legendre--Galerkin, Chebyshev--Tau spectral and Chebyshev--Collocation methods for experiment 2.}
\end{figure}

\begin{figure}[htbp]
\subfigure[]{\includegraphics[width=8cm]{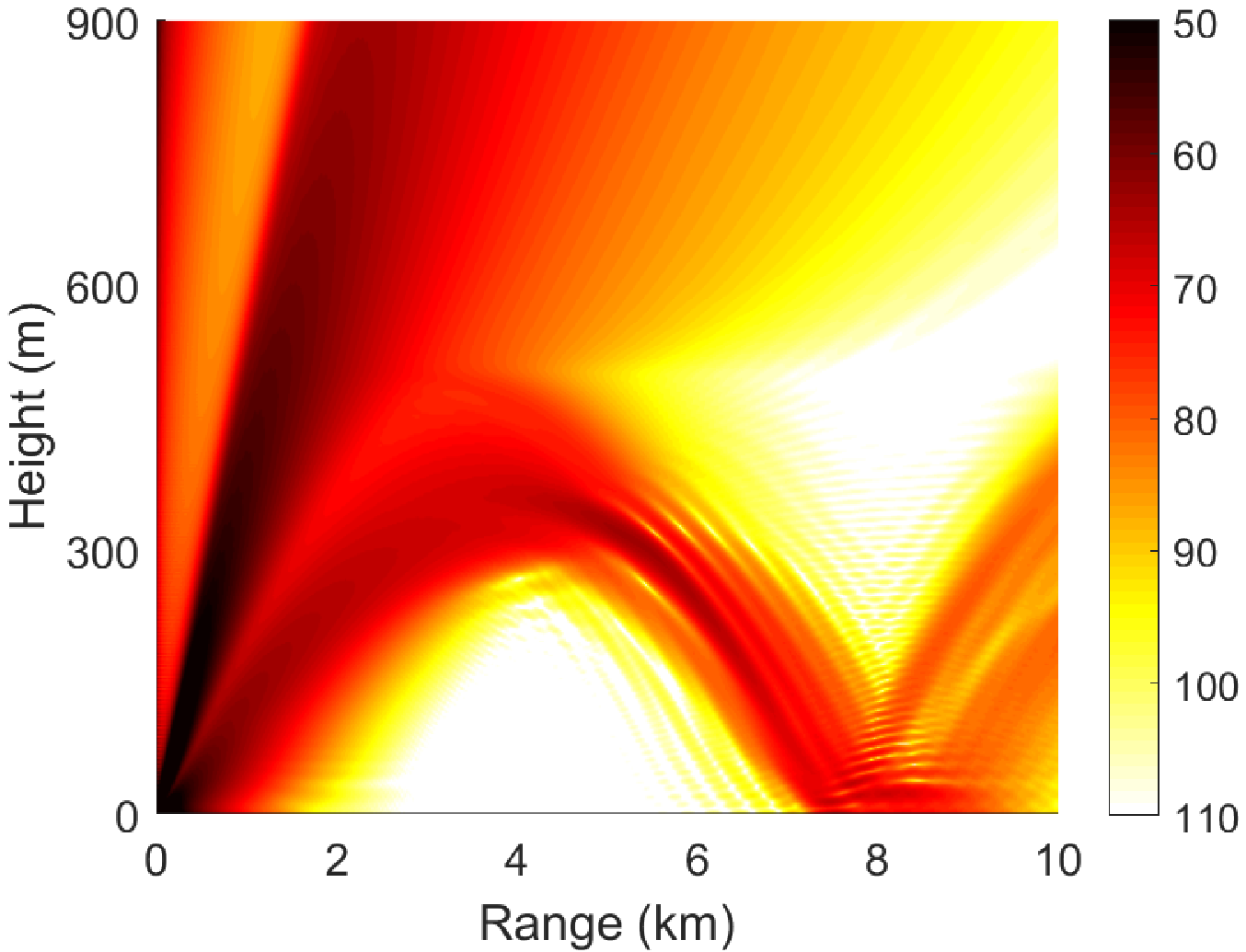}}
\subfigure[]{\includegraphics[width=8cm]{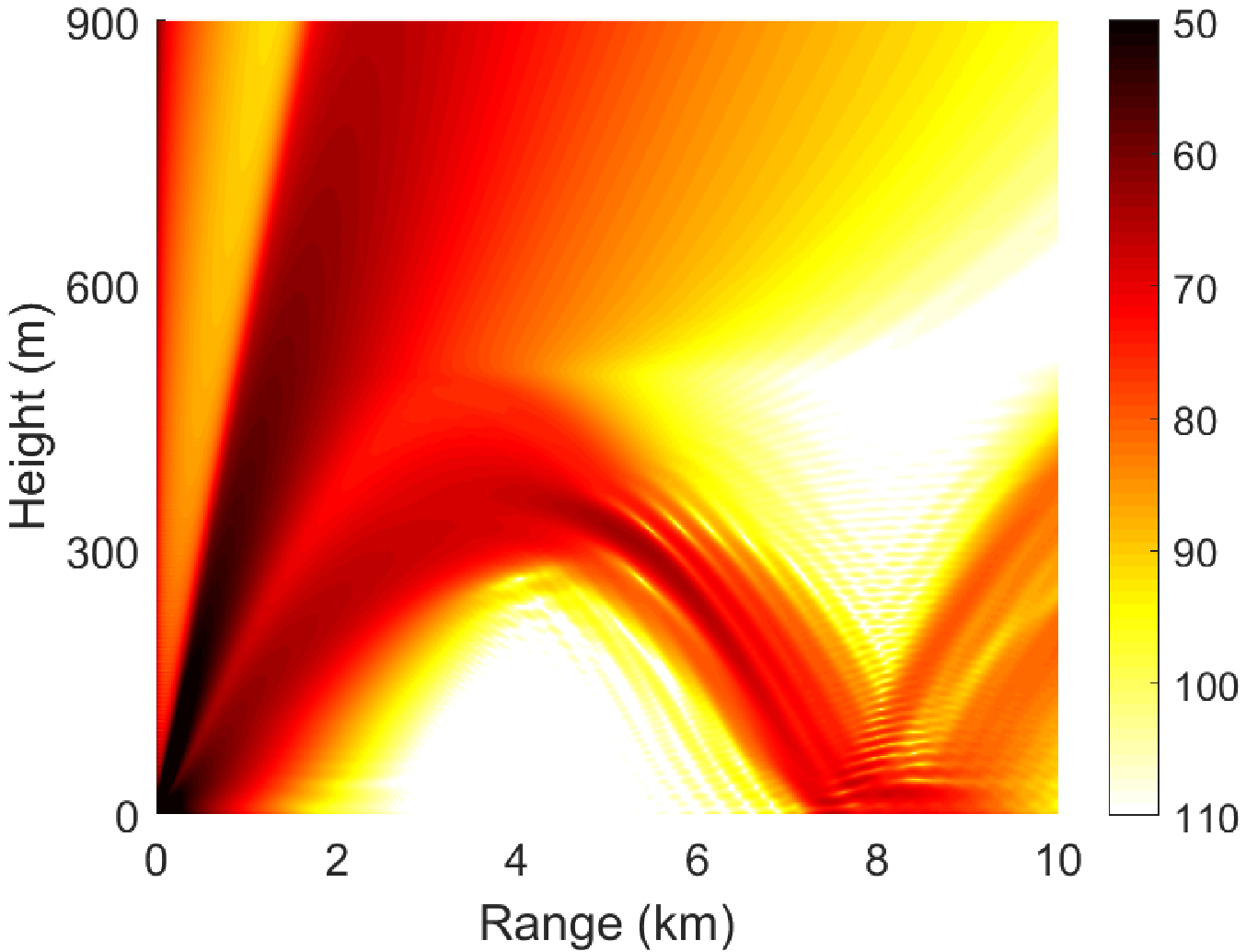}}
\subfigure[]{\includegraphics[width=8cm]{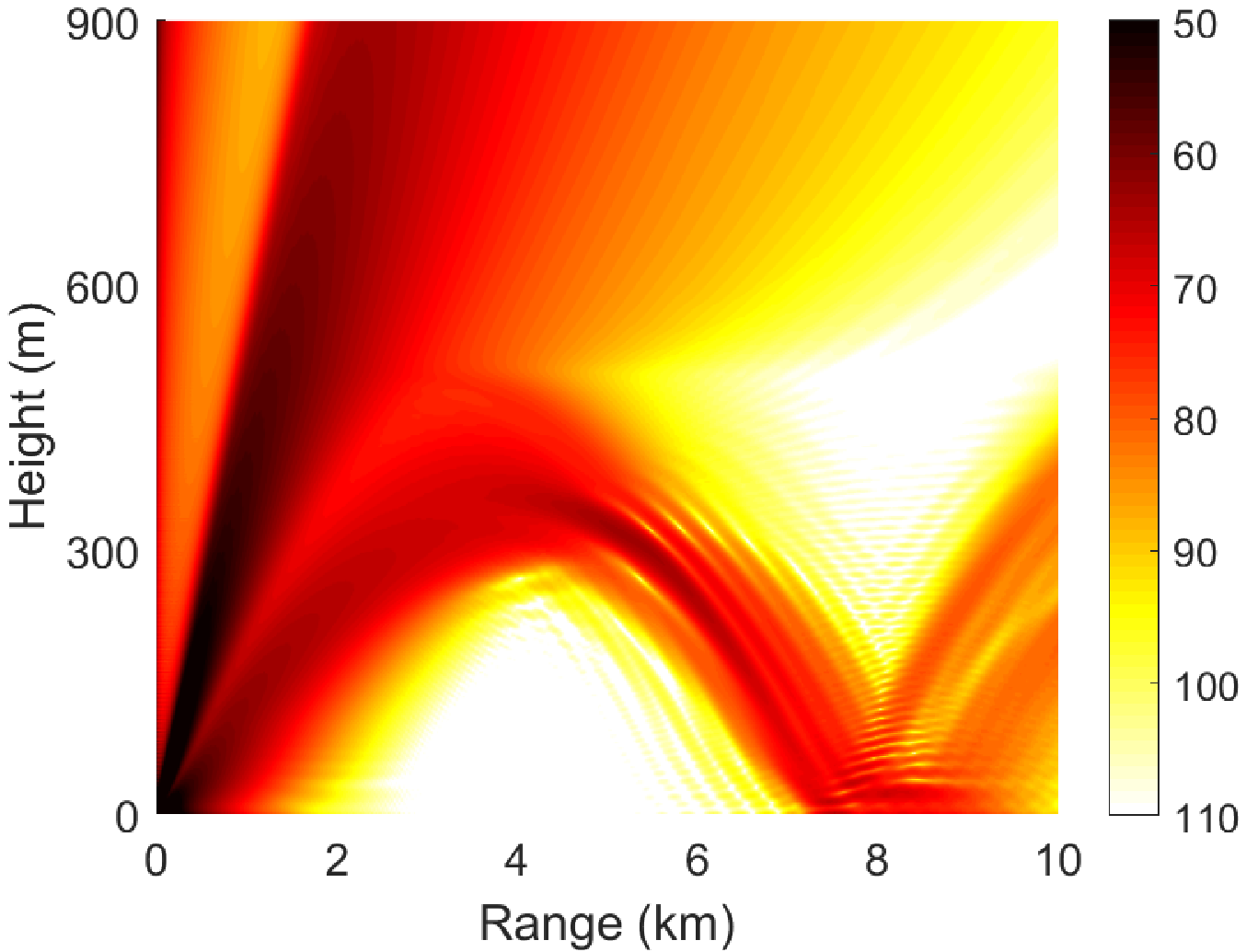}}
\caption{\label{fig7} Atmospheric acoustic fields obtained by (a) Legendre--Galerkin spectral method \cite{Evans2018}, (b) the Chebyshev--Tau spectral method, and (c) the Chebyshev--Collocation method for experiment 2.}
\end{figure}

\begin{figure}[htbp]
\includegraphics[width=8cm]{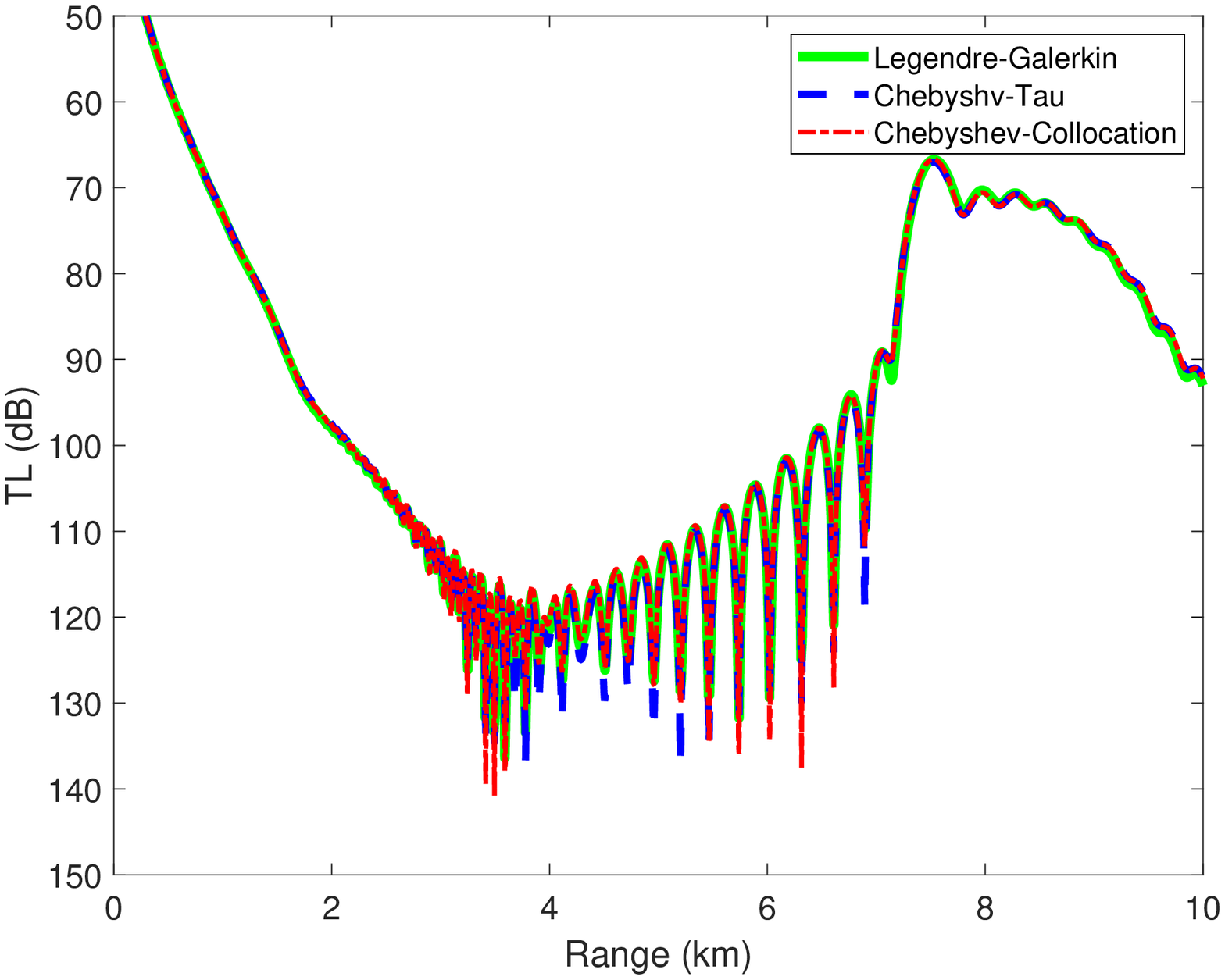}
\caption{\label{fig8} TL versus range for a receiver at a height of 1 m over the range interval 0--10 km from the (a) Legendre--Galerkin spectral method \cite{Evans2018} and the (b) Chebyshev--Tau spectral method and Chebyshev--Collocation method.}
\end{figure}

\autoref{fig6} shows four modes computed by the Legendre--Galerkin spectral method, Chebyshev--Tau spectral method, and Chebyshev--Collocation method. The modes obtained by the three methods are drawn in the same figure. The three lines almost completely overlap in the subfigures, and the differences between them are insignificant. \autoref{fig7} presents the acoustic fields calculated by the three spectral methods, where 553 modes with phase velocities less than 393.2 m/s were used to synthesize sound field. In the atmosphere layer, the sound fields calculated by the three methods were highly consistent. \autoref{fig8} shows the TL curves versus the range for a receiver at a height of 1 m over the range interval 0--10 km from the Legendre--Galerkin spectral method, the figure shows that the results of several methods were very similar. The differences between the three methods are indistinguishable at this plotting accuracy, and there may have been small differences only in the acoustic shadow area.

From the numerical results displayed above, we can see that three methods with different theoretical foundations all yielded very similar acoustic fields and normal modes, regardless of whether the sound speed profile was downwind or upwind. The consistency of these three methods proved that the two spectral methods proposed in this article are feasible for solving the atmospheric normal modes.

\section{Discussion of computational speed}
To further compare the characteristics of the two spectral methods proposed in this article, we divided each method into four steps, and we discuss the running time and complexity of each part separately. The four steps are as follows: discretizing the equation, solving eigenvalue problems, obtaining normal modes, and synthesizing the sound field. Table \ref{tab3} lists the time consumption of each step of the programs in the two experiments. The time listed in the table is the average of ten tests. In the tests, the three programs were run on a Dell XPS 8930 desktop computer equipped with an Intel i7-8700K CPU. The FORTRAN compiler used in the test was gfortran 7.5.0. 

In terms of speed, the AtmosCCSM was slightly faster than the AtmosCTSM. This is because the Tau method requires forward and backward Chebyshev transformations, unlike the Collocation method. The aaLG program was much slower than the two newly developed programs. Solving the eigenvalue problem was the most time-consuming step for the three programs. Moreover, the aaLG program spent much more time than the other two programs on solving the eigenvalue problem. However, the aaLG uses a subroutine developed by Evans \cite{Evans2018,Anderson1976} to solve the matrix eigenvalue problem, while both the AtmosCTSM and AtmosCCSM solve the eigenvalue problem by calling the Lapack numerical library. In fact, matrix eigenvalue problems have the same computational complexity $O(N^3)$. It is apparent from this table that the subroutine written by Evans is much slower than the Lapack numerical library, which is the main reason that the aaLG consumed much more time than the other two programs.

\begin{table}[htbp]
\tiny
\caption{\label{tab3} Time consumption of each step of each program in two experiments in units of seconds.}
\begin{ruledtabular}
\begin{tabular}{lccccc}
Experiment&Part of program &aaLG &aaLG-M &AtmosCTSM &AtmosCCSM \\
\hline
$\multirow{5}{*}{downwind}$
&1  &105.344 &104.691 & 0.522 & 0.468 \\
&2 &2017.324 &34.289 & 34.091& 34.331\\
&3     &35.587   &35.292 & 0.867 & 0.237 \\
&4   &10.021   &8.714 & 0.518 & 0.421 \\
&Total                      &2138.276 &182.986 & 35.998 &35.184\\
\hline
$\multirow{5}{*}{upwind}$
&1  &125.429 &123.892 & 0.482 & 0.361 \\
&2 &2039.324 &36.119 & 34.886& 34.017\\
&3     &36.501 &38.181  & 0.911 & 0.334 \\
&4   &11.669 &10.648  & 0.806 & 0.616 \\
&Total   &2212.923 &208.840 & 37.085&35.328 
\end{tabular}
\end{ruledtabular}
\end{table}

We modified aaLG to also call the Lapack numerical library when solving the eigenvalue problem. The modified program was named `aaLG-M'. The fourth column of Table \ref{tab3} lists the running time of aaLG-M. The aaLG-M took roughly the same time to solve the matrix eigenvalue problem as the other two programs. However, the aaLG-M was still slower than the other two programs. The most significant difference of the running time between the three programs was in the first step (discretizing the equation). In the first step, each element of the matrix finally obtained by the Legendre--Galerkin spectral method must be numerically integrated for every piecewise linear acoustic profile, which means the number of calculations is very large. In contrast, the two methods proposed in this article only need to perform simple interpolation of the acoustic profile and matrix multiplication to obtain the discrete equations. In the mode obtaining and normalization steps, the two spectral methods devised in this article still required less time than the Legendre--Galerkin method. It is worth mentioning that the AtmosCTSM.m and AtmosCCSM.m programs (developed in MATLAB, which is better at matrix operations) could obtain the results of the above experiments in less than 4 seconds (run on the same platform in MATLAB 2019a), which is an attractive result.

\section{Conclusion}
In this article, we propose two spectral methods for solving for atmospheric acoustic normal modes. An artificial absorption layer was added above the atmosphere of interest to reduce the impact of the truncated half-space on the area of interest. Next, we designed two examples, performed a detailed analysis of the results of each example, and finally verified the correctness and reliability of the proposed methods. Tests on the running time of the programs developed based on three methods showed that, in terms of the running time, the methods proposed in this article had better speeds than the Legendre--Galerkin spectral method.

\appendix*
\section{Chebyshev polynomials}
The Chebyshev polynomials are defined in the interval $x\in[-1,1]$ and have the following definition:
\begin{equation}
	\begin{split}
		&T_{0}(x)=1 ; \quad T_{1}(x)=x ; \quad T_{2}(x)=2 x^{2}-1,\\
		&T_{k+1}(x)=2 x T_{k}(x)-T_{k-1}(x),
		\quad k = 2, 3, 4, \cdots.
	\end{split}
\end{equation}
The orthogonality of these polynomials is defined as follows:
\begin{equation}
    \int_{-1}^{1} \frac{T_{j}(x) T_{k}(x)}{\sqrt{1-x^2}} \mathrm{d} x=\left\{\begin{array}{ll}
    0, & j \neq k \\
    \pi, & j=k=0 \\
    \pi / 2, & j=k \geq 1
    \end{array}\right.,
\end{equation}
where $\frac{1}{\sqrt{1-x^2}}$ is the orthogonal weighting factor. The expansion coefficients of a known function $\psi(x)$ on the Chebyshev polynomial basis can be obtained by the following formula:
\begin{equation}
\label{eq:a3}
    \hat{\psi}_{k}=\frac{2}{\pi c_{k}} \int_{-1}^{1} \frac{\psi(x) T_{k}(x)}{\sqrt{1-x^2}} \mathrm{d} x, \quad c_{k}=\left\{\begin{array}{ll}
    2, & k=0 \\
    1, & k>0
\end{array}\right..
\end{equation}
Eq.~\eqref{eq:a3} is called the forward Chebyshev transform, and it can be quickly calculated using the fast Fourier transform technique introduced by Canuto \cite{Canuto2006}.

\begin{acknowledgments}
The authors are very grateful to Richard B. Evans for providing the aaLG program in Ref.~\onlinecite{Evans2018}.

This study was supported by the National Key Research and Development Program of China [grant number 2016YFC1401800] and the National Natural Science Foundation of China [grant numbers 61972406, 51709267].
\end{acknowledgments}

\end{document}